\documentclass{ledger}
\usepackage{hyperref}

\usepackage{bbding}
\usepackage[]{algorithm2e} 
\usepackage{listings}
\usepackage{xcolor}
\usepackage{afterpage}

\usepackage{tablefootnote}
\usepackage{todonotes}
\usepackage{enumitem}
\usepackage{twemojis}
\usepackage{fontawesome}
\usepackage{pifont}
\newcommand{\xmark}{\scalebox{0.8}{\color{red} $\times$}}
\newcommand{\yesmark}{\scalebox{0.8}{\color{ForestGreen} \ding{52}}}

\newcommand{\thefirstpagenum}[0]{1}
\newcommand{\thelastpagenum}[0]{17}
\newcommand{\theyear}[0]{2024}
\newcommand{\thevol}[0]{X}
\newcommand{\thedoi}[0]{DOI 10.5195/LEDGER.\theyear.X}
\newcommand{\ledgerpages}[0]{\thefirstpagenum-\thelastpagenum}

\hypersetup{pdfauthor={Blake Regalia and Benjamin Adams}, pdftitle={Non-Fungible Programs}}

\title{Non-Fungible Programs: \\Private Full-Stack Applications for Web3}

\author{%
Blake Regalia\thanks{Solar Republic LLC, Washington, USA - \texttt{blake.regalia@gmail.com}}%
\and Benjamin Adams\thanks{University of Canterbury, Christchurch, New Zealand - \texttt{benjamin.adams@canterbury.ac.nz}}%
}

\pretitle{
  \centering \selectfont RESEARCH ARTICLE \par 
  \fontsize{24pt}{28pt}\selectfont} 

\begin{document}

\maketitle

\thispagestyle{pagefirst}

\begin{abstract}
The greatest advantage that Web3 applications offer over Web 2.0 is the evolution of the data access layer. Opaque, centralized services that compelled trust from users are replaced by trustless, decentralized systems of smart contracts. However, the public nature of blockchain-based databases, on which smart contracts transact, has typically presented a challenge for applications that depend on data privacy or that rely on participants having incomplete information. This has changed with the introduction of confidential smart contract networks that encrypt the memory state of active contracts as well as their databases stored on-chain. With confidentiality, contracts can more readily implement novel interaction mechanisms that were previously infeasible. 
Meanwhile, in both Web 2.0 and Web3 applications the user interface continues to play a crucial role in translating user intent into actionable requests. In many cases, developers have shifted intelligence and autonomy into the client-side, leveraging Web technologies for compute, graphics, and networking. Web3's reliance on such frontends has revealed a pain point though, namely that decentralized applications are not accessible to end users without a persistent host serving the application. Here we introduce the Non-Fungible Program (NFP) model for developing self-contained frontend applications that are distributed via blockchain, powered by Web technology, and backed by private databases persisted in encrypted smart contracts. Access to frontend code, as well as backend services, is controlled and guaranteed by smart contracts according to the NFT ownership model, eliminating the need for a separate host. By extension, NFP applications bring interactivity to token owners and enable new functionalities, such as authorization mechanisms for oracles, supplementary Web services, and overlay networks in a secure manner. In addition to releasing an open-source software development kit for building NFPs, we demonstrate the utility of NFPs with an interactive Bayesian game implemented on Secret Network.

\end{abstract}

\section{Introduction \& Motivation}

In the decade since the development of the Ethereum network, the promise of decentralized applications powered by smart contracts has been touted as an integral part of the vision of Web3 \cite{buterin2014next}. The vision idealizes myriad decentralized applications (dApps) spanning finance, gaming, healthcare, real estate, energy, censorship-resistant social media, and the metaverse. In contrast to this aspirational and wide-ranging vision, the most notable real-world successes for smart contract-enabled applications have focused on decentralized finance (DeFi), where self-custody and trading of fungible cryptocurrency tokens can happen without the role of an intermediary bank \cite{zetzsche2020decentralized}. The recent boom (and subsequent bust) in markets for non-fungible tokens (NFTs) used as signifiers of ownership for digital art, mark a more mixed result for the technology \cite{wang2021non,hofstetter2022crypto}. On the one hand it demonstrated that smart contracts can have utility for applications with broad appeal beyond finance. However, the phenomenon of ``right click save'' clearly showed that value comes not only from a digital certificate of ownership but also possession of the content of the owned thing as well. 

Beyond these applications, the promise of Web3 has largely been unrealized. The user experience, adoption, and functionality of decentralized versions of applications often compare poorly to more centralized Web 2.0 applications \cite{murray2023promise}. Arguably the greatest successes of Web3 have been in new types of applications outright (such as DeFi), which did not exist previously, more so than refactoring existing Web applications to run on the blockchain. This is because the added value of self-custody and decentralized automation that blockchains and smart contracts provide rarely overcomes deficits in usability, especially when compared to web applications that build upon a suite of mature technologies to achieve heavy computation, high-end graphics processing, privacy, secure transmission of data, and responsive interactivity. 

Confidential smart contract blockchain networks encrypt contract memory and data stored on chain allowing us to build new types of dApps that are greater than the sum of their Web 2.0 or Web3 parts \cite{secret,oasis}. 
In Web 2.0, transport layer security (TLS) is an integral technology for applications because it encrypts transmitted data, adding trust for users and facilitating regulatory compliance. Without this level of security and privacy much of the modern Web would not exist. 
Still, many Web applications have a shared global state stored on a centralized server, which relies on a trusted administrator. 
Confidential smart contracts align smart contracts with technologies like TLS because the computation on-chain is privacy-preserving providing similar guarantees for on-chain computation that TLS does for off-chain communication. This adds a layer of security and privacy to decentralized applications that require trustless automation. However, although applications that combine a Web frontend with confidential smart contracts exist, the frontends are invariably hosted on a centralized server.

The solution presented in this paper is to introduce the concept of the Non-Fungible Program (NFP), an extension of the non-fungible token (NFT).
An NFP represents a privately-held token (NFT) which grants its owner exclusive access to a hidden-state, decentralized application that uses smart contracts for its backend and self-contained Web documents for its frontend. Key contributions presented in this paper include:

\begin{itemize}
    \item A method for encapsulating self-contained applications using SVG documents that transform into HTML5 web applications with exclusive access to private modules and assets stored in a confidential backend smart contract.
    \item An on-chain package manager system for executable code that hosts immutable and always-accessible versions.
    \item A software development kit for NFPs and demonstration running on Secret Network.
\end{itemize}


\section{Related Work \& State of the art}

\subsection{Progressive Web Apps (PWAs)}

Progressive Web Apps (PWAs) were coined as a marketing concept but have coalesced around a set of common Web technologies that allow applications to implement features for consistent user experience \cite{hume2017progressive}. The growth of PWAs came out of a desire to make Web applications on mobile platforms match the features of native applications by being installable and runnable offline \cite{biorn2017progressive}. However, in contrast to native applications, PWAs are also an approach for cross-platform development, because browsers that can run Web applications are ubiquitous on network-enabled devices. Other key features of PWAs include push notifications and background synchronization. PWAs are made possible by two key Web technologies: Service Worker support in browsers and HTTPS \cite{steiner2018web}. Service Workers operate as proxies that can choose when to serve cached data instead of fetching from a remote host when completing a web request. Browsers require that Service Workers run in a Secure Context \cite{West:21:SC}, which HTTPS satisfies. An application's source code and data are therefore able to function even if the client doesn't have an active internet connection, i.e., once the PWA has been installed.

\subsection{Privacy for blockchains}

Addressing privacy in blockchain networks is an active research area. Non-interactive zero-knowledge proof (ZKP) systems, such as Succinct Non-Interactive Argument of Knowledge (ZK-SNARK), Scalable Transparent Argument of Knowledge (ZK-STARK), and Bulletproof have become popular tools for building privacy mechanisms into networks \cite{ben2014succinct,ben2018scalable,bunz2018bulletproofs}. In blockchains, zero-knowledge proof systems can probabilistically check that statements are true about transactions and balances, and have become integral features of privacy cryptocurrencies \cite{sun2021survey}. In smart contract networks, they can also be used for authentication and identity management, and to verify that a computation has occurred, opening up the opportunity to move expensive but verified computation off-chain \cite{yang2020zero,buterin2021incomplete}. Since many dApps also depend on oracles to provide off-chain data to smart contracts, zero-knowledge proofs can also prove the authenticity of a data source while preserving privacy \cite{zhang2020deco}. Although ZKPs are a useful mechanism to provide many privacy features for users of blockchains, they are only suitable for a specific form of problems between a prover and verifier. Other classes of problems involving private data from multiple sources---e.g., a shared program memory state contributed to by multiple users each with partial knowledge---require different solutions. 

Confidential smart contract networks approach the problem of privacy in blockchains in a complementary but fundamentally different way. Confidential smart contracts are general purpose blockchain programs that enable some form of default privacy for contracts as they run (in contrast to network-level transactional privacy) \cite{zyskind2015decentralizing,kosba2016hawk,cheng2019ekiden}. Active programs do not reveal internal state to an outside observer, including an administrator of the node executing the code, and the data stored on chain is encrypted in a format only readable by the contract. Currently, the most practical implementations of confidential smart contracts rely on the use of hardware secure-enclave encryption to achieve confidentiality. Two networks---Secret Network\cite{secret,woetzel2016secret} and Oasis Network\cite{oasis}---have working mainnets using this approach. Secret Network is a Tendermint-based network using a heavily-modified version of the Cosmos Internet of Blockchains software development kit \cite{buchman2016tendermint}. Smart contracts on Secret Network are written in Rust and stored as Web Assembly (WASM) binaries. Oasis Network uses a parachain model with public and private networks built on an extension of the Ethereum Virtual Machine (EVM). Although the networks differ in many ways, both networks use Intel Software Guard Extensions (SGX) to execute contract code inside a secure trusted execution environment (TEE) for private computation \cite{costan2016intel}. Reliance on hardware for confidentiality is not without risks, as exploits can arise due to the setup of secure enclave keys, side-channel attacks, patching regimes for nodes on the network, and access pattern leakage \cite{jean2023sgxonerated}. 

Confidential dApps vary in terms of long-term and short-term privacy obligations \cite{casassa2004dealing}. For example, in a game of incomplete information between two or more players, the imperative of confidentiality only lasts as long as the game is being played, so as not to benefit any player. In contrast, the medical history of an individual would need long-term privacy guarantees. Beyond hardware-based encryption, there is current research to implement general-purpose confidential smart contracts using fully-homomorphic encryption (FHE), which would eliminate such risks \cite{gentry2009fully}. The NFP model we present here is not dependent on any one type of confidential smart contract implementation. However, hardware confidentiality is likely to remain the most viable option for developing NFPs in the near term, because even as these new methods are developed, they will be far less computationally-efficient than TEEs.

\subsection{Non-fungible tokens}\label{sec:nfts}

A non-fungible token (NFT) is a digital representation of ownership over a unique asset recorded on a blockchain \cite{wang2021non}. The rules for minting, transferring ownership, and authenticating NFTs are governed by smart contracts. Non-fungible token standards, such as ERC-721 on the Ethereum network, are used to define common behavior for NFTs enabling them to be bought and sold on marketplaces \cite{erc721}. Digital content that an NFT refers to can be stored entirely on chain, but commonly only metadata is stored on chain and will reference content stored off chain. In the case of digital art NFTs this external content is often stored on peer-to-peer data networks, such as the InterPlanetary File System (IPFS), in an effort to decentralize distribution. All aspects of NFTs stored on public blockchains are publicly readable, including full ownership history as well as any associated content stored off chain.

Secret NFTs are NFTs that are created using the SNIP-721 standard on Secret Network \cite{snip721}. Other confidential smart contract networks such as Oasis Network also have the capacity to implement Secret NFTs, though no equivalent standard exists, yet. Because token data and metadata are encrypted, owners of Secret NFTs are able to irrevocably maintain exclusive access to digital content and prove ownership 
in zero-knowledge exchanges. 


\subsection{Web application categories}

Analogous to the enabling role that Service Workers and HTTPS play for PWAs, confidential smart contracts are foundational for implementing NFPs. We compare Web applications across four qualifiers as follows:
\begin{itemize}
  \item Decentralized -- Is the backend service neither operated nor controlled by single authority?
  \item Private -- Is the backend storage of user data provably private?
  \item Hostless -- Does access to the content/application critically rely upon a connection to an active Web provider? Or is the content accessible without an active Web provider?
  \item Computable -- Is the frontend capable of client-side computation?
\end{itemize}

Table~\ref{tab:webapps} compares the various Web application types using these qualifiers. Single-page applications (SPAs) and PWAs are indicative of Web 2.0's capacity to provide computable functionality on the Web. DeFi, Secret DeFi (DeFi with confidential smart contracts), NFTs, and Secret NFTs are Web3 artifacts that vary in terms of hostless-ness and frontend computability. The NFP model we propose in the following section satisfies all four qualifiers.

\begin{table}[ht]
\centering
\begin{tabular}{lllll}
 & Decentralized & Private & Hostless & Computable \\
SPA
& \xmark   & \xmark   & \xmark   & \yesmark \\
PWA
& \xmark   & \xmark   & \yesmark & \yesmark \\
DeFi
& \yesmark & \xmark   & \xmark   & \yesmark \\
Secret DeFi    & \yesmark & \yesmark & \xmark   & \yesmark \\
NFT            & \yesmark & \xmark   & \yesmark & \xmark   \\
Secret NFT     & \yesmark & \yesmark & \yesmark & \xmark   \\
NFP            & \yesmark & \yesmark & \yesmark & \yesmark           
\end{tabular}
\caption{\label{tab:webapps}Comparison of application types on the Web.}
\end{table}

\section{Non-Fungible Program Model}\label{methods}





\begin{figure}[!htp]
  \centering
  \includegraphics[width=1\linewidth]{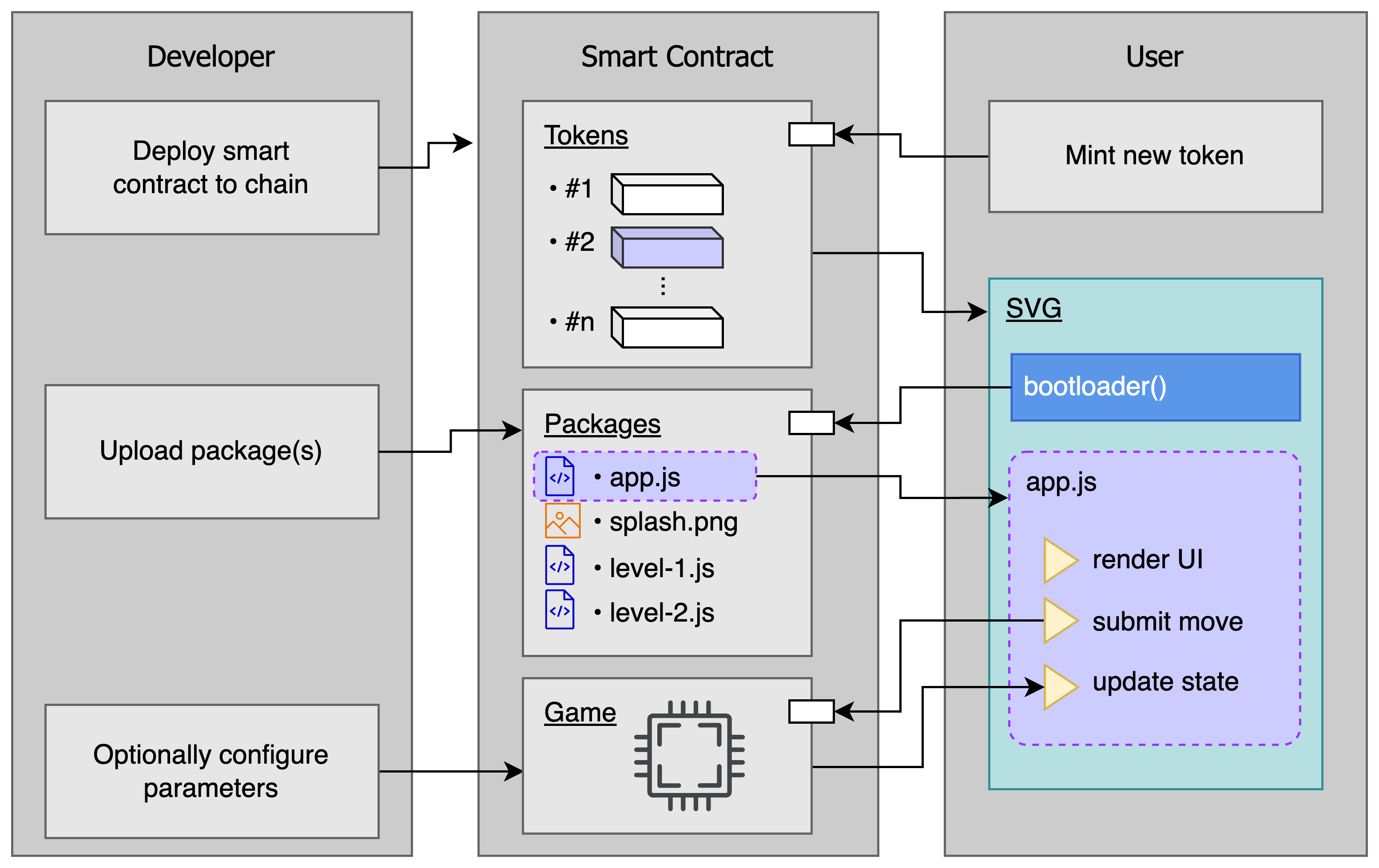}
  \caption{Schematic diagram of a sample NFP game deployment and its execution, which shows the interactions between the developer, the on-chain smart contract, and the user.}
  \label{fig:concept-diagram}
\end{figure}

In this section we detail the NFP model, covering the methods involved in creating a decentralized, private data service backend, the computable Web application frontend that runs in a browser, and the interface to couple them together. As a reference for the following section we provide a concept diagram illustrating some of the key elements of the model and their relationships for a sample NFP deployment in Figure~\ref{fig:concept-diagram}.

\subsection{Ownership}

Unlike other digital assets which can be replicated without limitation, one of the fundamental tenets of NFTs is that they offer verified proof of authenticity and proof of ownership over a particular digital item or piece of content. Moreover, an NFT collection can be designed to allow for ownership to be transferred, or forbid it entirely. Since transferring an NFT executes its contract, collections will often enforce custom rules, restrictions or behaviors on transfer events such as minimum qualifications, trading windows, or royalties, respectively. These extensible regulations on ownership are what allow NFTs to serve such a wide variety of use-cases. As described in Section~\ref{sec:nfts}, Secret NFTs built using confidential smart contracts enable exclusive access to content. NFPs adopt the Secret NFT ownership model.

Ownership is enforced by the NFP contract. New NFPs can be `minted' by authorized minter accounts executing a mint transaction, the specifics of which depend on the application. As NFPs are minted, an encrypted index of owners is maintained. The contract provides logic to transfer token ownership and to enforce access permissions to view private, exclusive data. For Secret NFTs, data can be stored entirely on-chain or in the case of large files, a secret key can be stored on-chain to provide access to off-chain encrypted data. With NFPs, storage of the main Web document and associated packages are stored fully on-chain (details below).



\subsection{Backend \& Infrastructure}


NFPs critically rely on the concept of confidential smart contracts, which by design are immutable
and keep their active memory state hidden. This enables the development of trustless backend services that can execute logic based on a global encrypted state. Smart contract code is executed and new blocks are proposed by nodes within the blockchain network. A full exploration into blockchain network architectures is beyond the scope of this article. However, both Secret Network and Oasis Network have a set of validator nodes that use a Proof-of-Stake consensus mechanism to verify new blocks written to the chain. Public API endpoints exist to receive encrypted, signed transactions from external applications and return responses, and anyone can upload and interact with contracts on the chain. NFPs rely on such an infrastructure to provide decentralization. All of our experiments using the NFP model have been tested on the Secret Network testnet (Pulsar-3) infrastructure, but future work will expand to other networks.

\subsection{SVG}

Scalable Vector Graphics (SVG) is a broadly-supported, versatile XML-based file format developed by W3C that encapsulates a broad spectrum of visual representations \cite{Lilley:11:SVG,Jackson:03:SVG}. In its simplest form, an SVG document can be rendered as a static image scalable to any resolution. The static image representation is most commonly seen in the thumbnail preview of file browsers or image galleries, or when used as an icon in application launchers. We refer to this most basic viewing mode as the “First Frame Preview” of an NFP, acting as a splash page to the interactive content within. Ideally, the preview also conveys what unique characteristics of the token make it non-fungible, in the same manner that JPEGs do for classic NFT collections.

The Synchronized Multimedia Integration Language (SMIL) is a W3C markup language capable of adding transitions, animations, and a limited range of interactivity to SVG documents \cite{Hoschka:98:SMI}. Some operating systems’ default image previewer will play SMIL animations for SVG documents, but disable CSS animations and prevent scripts from executing. Similarly, SMIL animations are enabled when an SVG is linked or embedded in an HTML document no matter how it is referenced, although SMIL interactivity will differ depending on the embedding method\footnote{\url{https://www.w3.org/TR/SVG2/conform.html\#examples}}. Given these capabilities and constraints, developers are free to embellish their SVG with dynamic content in what we refer to as the “Active Preview” viewing mode. For example, a collection of mythical creature NFPs might give each character a breathing animation in order to evoke a sense of liveliness. In some contexts, attribute changes and animations can be triggered in response to input events, such as mouse clicks, which allows for some basic interactivity. For example, clicking on a character’s head might trigger an emote and reveal a speech bubble, adding to the interactive depth of the preview.

SVGs opened directly in a modern web browser, or embedded in an HTML document without sandboxing, typically have the ability to execute scripts. With scripting enabled, the SVG is able access the DOM and available Web APIs including Fetch, Crypto, WebRTC, and WebGL, to name a few. By appending an SVG \texttt{foreignObject} element to the document once opened, the web page can transform into a modern HTML5 web application.

When a downloaded NFP SVG document is opened directly in the browser, it is served from the \texttt{file:} scheme, imposing certain security restrictions. As the Secure Contexts specification advises, ``the user agent SHOULD treat file URLs as potentially trustworthy'' \cite{West:21:SC}, meaning that implementations may vary. This could affect an application’s ability to leverage some advanced or experimental Web APIs including but not limited to Service Workers, Push, Geolocation, Web Authentication, Web USB, and Web Bluetooth. However, the breadth of features available to an application served from the \texttt{file:} scheme is powerful enough for most use cases we envision, including online gaming, social media, and marketplaces for digital content.

If an application depends on a Web API that requires a Secure Context, or if a user wants to audit or control the requests being made by an SVG, then a special sandboxing web application served over HTTPS can be used to host the SVG as it runs. In this paper, we do not evaluate sandboxing but plan to build such an application in future work.

\subsection{Reliability}

Running a Secret Network node requires SGX hardware to create Trusted Execution Environments. A collection of public API endpoints are hosted and funded by the community. Embedded into each token's SVG are a set of URLs pointing to API endpoints so that the frontend can communicate with the blockchain. Since the SVG stored on-chain is immutable, there is a risk of the embedded URLs becoming obsolete over the long-term. However, clients are free to locally modify an SVG's contents before launching it without compromising the application. In order to ensure these URLs are accessible for change, we introduce an XML namespace for NFPs and embed a set of configurable metadata properties in the SVG document in the form \texttt{<metadata><nfp:web nfp:lcds="..."/>}. These XML elements are designed to be both human-readable and machine-readable, the latter being useful for NFP sandboxes.

\subsection{Packages}\label{packages}

An NFP's SVG is a self-contained document; it does not use any native import methods to include assets from the web. All graphics, styling, and scripts are embedded directly into the SVG file and its contents get permanently written to the blockchain upon token mint. The on-chain contents are immutable by design. However, immutability poses a challenge to maintainability. NFP publishers may discover bugs in their application or may wish to add new features after launching a production mint. To address this shortcoming, we introduce the concept of a package manager as part of the NFP contract interface specification. With packages, admins of an NFP are able to publish revisions to parts of their application.

Analogous to traditional package manager systems, a \textit{package} refers to a collection of \textit{package versions}, where each version has its own data payload and metadata. Each package must be given an identifier that is unique within the contract. The identifiers do not need to be opaque nor human-readable since it is the responsibility of the developer to manage a naming scheme for their NFP. For example, a simple convention would be to use the basename of the file with its extension for readability, e.g., ``main.js'', ``sprites.webp'', ``models.gltf''. A more advanced system on the other hand might benefit from namespacing each package to avoid conflicts between manually uploaded and contract-generated packages.

In order to help guarantee token owners irrevocable access to application features, the contract code for the package manager ensures that all versions of a package are immutable and remain forever accessible to the owners of tokens for which they are designated. In our draft specification, we define four distinct \textit{access specifiers} that determines how access-control is applied to a package. Once created, a package’s access specifier cannot be changed in future versions. Keep in mind that \textit{package} refers to a series of package versions immutably stored on-chain. We describe each of the four access specifiers below:

\begin{description}
\item[\texttt{public}] -- No authentication is required. Any client can anonymously query the contract to access the package.

\item[\texttt{owners}] -- Only accounts owning at least one token are authorized to view the package.

\item[\texttt{cleared}] -- Token owners must be individually approved to access the specific package. For example, a game may require the user to clear level 1 and prove it to the contract before they are allowed to fetch the module for level 2. Implementors are free to choose whether transferring an NFP resets a token’s cleared status on a per-package basis.

\item[\texttt{token}] -- The package was specifically created for an individual token, and that token's owner has exclusive access to the package. Whereas admins are able to publish and view packages having the 
other specifiers,
they are explicitly forbidden from publishing or viewing packages with the \texttt{token} specifier. Consequently, the only means to publish a package having the \texttt{token} specifier is through contract logic. For example, a contract might generate a package for a specific token based on some rules and an internal secret. Another use case might be to provide token owners a private, decentralized, version-controlled file system.

\end{description}

When a client needs to retrieve a package's contents, it might not be desirable to always return the most recent version. There are a variety of use-cases that depend on multiple versions of a package being accessible simultaneously. We found that using a tagging scheme provides the most flexibility for these use cases. When uploaded, each package version can be assigned an optional set of tags, e.g., ``latest'', ``1.x'', ``beta'', and so on. When clients query for a package version, they can either specify the exact version by its serial number, or select the most recent version that includes a given tag, such as ``latest''.

\subsection{Storage}

Storage is a limited resource for decentralized applications. Most NFTs use third-party overlay networks such as the InterPlanetary File System (IPFS) to store the contents of digital assets \cite{benet2014ipfs}. Some Web3 applications have experimented with using IPFS to host their client-side code. However, IPFS also brings its own set of challenges when it comes to data longevity and accessibility \cite{daniel2022ipfs,doan2022towards}. For example, each file requires persistent seeds in order to be available on the network. Other factors such as peering and network topology can also affect clients differently, playing an important role in download bandwidth. As for hosting application frontends, current solutions require special software or augmented browsers in order to handle routing requests for lookups such as the Ethereum Name Service.

With our NFP approach, the application's frontend code is considered part of the token's digital asset, and it is privately stored directly on the same blockchain as the token. This approach eliminates the dependency on a third-party content hosting solution while preserving all the benefits of an immutable, decentralized storage provider. Additionally, it secures the contents automatically, i.e., without requiring a separate encryption layer. Secret Network was built using the Cosmos SDK\cite{cosmossdk} and extends the CosmWasm\cite{cosmwasm} smart contract platform, which uses a floating key-value store to persist contract state. For CosmWasm-enabled chains, the `at-the-pump' gas cost per byte written to storage from a contract depends on several factors including current market value of the gas token and certain network parameters. Similarly, the maximum size of a bytestream that can be uploaded to a contract's storage in a single transaction depends on network parameters and can be affected by validator configurations. As we will discuss in Section~\ref{evaluation}, the effective costs and limitations observed on Secret Network currently make our storage solution feasible, but may impose greater challenges if network consensus were to change certain parameters. It may be the case that storing large assets in contract memory is prohibitive on other privacy chains. A possible workaround to upload size limitation we have explored involves chunking an asset into multiple transactions and then reassembling upon retrieval.

Given that storage in contract memory is a valuable resource for decentralized applications, each package version contains an attribute for content encoding, e.g., \texttt{gzip}. Consequently, the host environment running the package code should support the ability to decode compressed formats. Fortunately, the Compression Streams Web API provides this functionality natively in the browser.

\subsection{Bootloading}

\newcommand{\Client}{Client}

Once an SVG is opened in a browser with the ability to run scripts, it transforms into an HTML5 web application which we refer to as the \textit{\Client{}}. The transformation has several discrete steps that we will describe in this subsection.

In all cases, an NFP \Client{} will need to retrieve private data from the contract. To do that, it must encode RPC requests and submit them over the HTTPS transport. Querying and executing a confidential contract on the Secret Network requires encrypting the contents of the message that will get sent into the node's secure enclave where it will then be decrypted and forwarded to the target contract. Similarly, the \Client{} must also decrypt the response data in order to read the results returned by the contract. This process requires several cryptographic functions that are not available in the Web Crypto API. 

While the entire \Client{}'s source could be contained in the SVG file, a more flexible design pattern, as we mentioned in Section \ref{packages}, is to load the application in modules by leveraging the contract's package manager. With this approach, each token's SVG can query the chain for the ``latest'' stable version of the application's entrypoint JavaScript module and evaluate it by appending a \texttt{script} element to the DOM. We refer to this process as \textit{bootloading}, and the requisite script embedded in the SVG as the \textit{bootloader}.

Using a bootloader and modules to defer loading the application this way also helps reduce the size of each token's SVG since the same module can apply to multiple tokens. A reduction in the size of the SVG ultimately has cost savings in terms of gas paid by users during mint operations. One trade-off to this approach is that it may take a few seconds longer to start up when the SVG is opened or reloaded as the bootloader has to wait for the query response from the network. A possible mitigation however is to cache the module locally upon initial download and then check if a newer package version is available in the background on subsequent loads.

\subsection{Querying and Executing}\label{queryingexecuting}

With the NFP contract acting as a backend service to the \Client{}, there are two types of requests that can be made, queries and transactions. Queries are the means to call functions on the contract that do not write data to the blockchain, but are still able to read from the contract’s private database and perform confidential computing. On Secret Network, the query request and response are always encrypted, but clients must also authenticate with the contract to prove they have permission to view the requested information. Contracts may expose both public (un-authenticated) and private (authenticated) query methods to clients. One popular authentication technique involves the account owner digitally signing a document known as a Query Permit. A signed Query Permit grants any client with it in their possession the ability to perform some specified set of read-only queries on behalf of the signer. For example, Alice signs a Query Permit designating contract X and the ability to perform all private queries. She then transfers the signed Query Permit to Bob who submits it with his query to contract X asking for the token balance of Alice’s account. Contract X verifies the signature on the document using Alice’s public key, checks the permissions granted, and returns Alice’s token balance. As long as the authentication passes, whether Alice or someone else submitted the query is inconsequential to the query process. Query Permits can later be revoked by executing a transaction.

Executions are used to mutate the contract’s internal state. An execution is carried out by writing a transaction to the blockchain. Like queries, executions are also able to read from the contract’s private database and perform confidential computing. Unlike queries, executions are automatically authenticated since each transaction message requires the account owner’s digital signature. Whereas queries can be served by the secure enclave within any compliant, participating node on the network, executions can only be performed by validator nodes securing the network through proof-of-stake. The amount of confidential computing time an individual query or execution may consume is regulated in terms of gas consumption, but the gas used for an execution must be paid for while the gas used for queries do not.

For supporting the combination of queries and executions, the \Client{} must perform cryptographic operations that are not natively supported by the browser or any Web API. At minimum this includes Bech32 encoding/decoding \cite{wuille2017bip173}, elliptic curve scalar multiplication on Curve25519 \cite{bernstein2006curve25519}, RIPEMD-160 hashing \cite{dobbertin1996ripemd}, AES-GCM-SIV \cite{rfc8452}, Secp256k1 \cite{brown2010sec2} key generation, ECDSA \cite{johnson2001elliptic} (elliptic curve digital signature algorithm) for signing/verification and ECDH (elliptic curve Diffie-Hellman) for asymmetric encryption.
As we discuss in Section \ref{sdk}, we publish a runtime library for NFP \Client{} developers covering all the aforementioned operations, including common abstractions for interacting with Secret Contracts and other modules on the Secret Network.

\subsection{Fee Grant and Delegation}

When executing contracts from the \Client{}, a potential point of friction for the user experience in NFPs is the frequency of transactions that must be signed by a compatible wallet representing the token owner's account. For example, every move in a game of Chess might require the player to sign a transaction on their air-gapped hardware wallet, making the experience inconvenient enough to avoid entirely. Here we present a solution that combines a native network feature with smart contract authorization by creating a `hot' hot wallet in the browser to execute transactions on behalf of the token owner without prompts.

Cosmos SDK provides a mechanism called Fee Grants that allows users to pay their transaction fees from the balance of another account. The granter signs and broadcasts a transaction approving some grantee with an optional spending limit and expiration. When the \Client{} loads, it generates a securely random private key for a new hot hot wallet account and saves it to local storage. It then requests a Fee Grant approval to grant the new account some limited spending allowance, ensuring that the actual balance of the new account remains nil in case it is compromised. The spending limit ensures that the entire balance of the granter isn't at risk of being wasted on transaction fees.

The hot hot wallet account also needs authorization to execute contract methods on behalf of the token owner. Rather than granting the new account unlimited discretionary power, our approach with the NFP contract interface specification defines a set of operations to approve or revoke delegate accounts on a per-owner or per-token basis. For example, the \Client{} for an NFP chess game might request token delegate approval to execute only a limited subset of methods, such as creating new games, joining existing games, and submitting moves.

In summary, Fee Grant and execution delegation provides reasonable security such that if the hot hot wallet account of the \Client{} were to be compromised, an attacker would not be able to transfer ownership, steal funds, or drain the owner's account balance.

\subsection{Supporting Development}\label{sdk}

An important consideration for NFPs as a platform is the developer experience. Streamlining the build process, making reusable components, and defining interfaces are crucial elements to fostering resources for developers. As part of our work, we have published open-source tooling and draft specifications to make these resources available to developers from the start. Namely, we define a contract interface specification for NFP contracts that includes query and execution methods for ownership, transfer-ability, delegation, package management, private notifications, and key-value storage. On the \Client{}-side, we define an XML namespace and element schema to provide SVGs with human-readable and machine-readable metadata, including configuration for API endpoints.

For the development of scripts for the \Client{}, a custom NFP SDK provides a suite of tools designed to automate best practices and parts of the build process. For example, the SDK exports a Vite plugin allowing developers to use ES import and export syntax to create pseudo-modules that work across package boundaries. By default, the plugin also strongly favors producing small distributables to be uploaded to the chain by bundling and minifying scripts and optimizing the SVG. Finally, as mentioned in \ref{queryingexecuting}, an NFP runtime library provides all the functionality needed to transact with Secret Contracts, including bootloading and private notifications.









\section{Evaluation}\label{evaluation}

We evaluate our approach by implementing a two-player, turn-based Bayesian game as an NFP \cite{mertens1985formulation}. 



\subsection{The Game}

Inspired by the classic game Salvo and derivatives such as Battleship, two players compete in a zero-sum game by taking turns submitting a coordinate to attack on their opponent's 10x10 grid. At the beginning of the game, each player configures five stationary `vehicles' of varying lengths to occupy the available cells of their own grid which is hidden from their opponent. The objective is to be the first to destroy all five of their opponent's vehicles. Each vehicle can be placed on the grid either horizontally or vertically, must be entirely contained within the grid, and cannot overlap with other vehicles. 
After submitting each attack, the player is informed whether or not it struck a cell occupied by a vehicle, i.e., either hit or miss. Once a player has completely destroyed a vehicle by striking all of its cells, the vehicle's type and the coordinates it occupied are revealed to the player.

\subsection{Requirements}\label{requirements}

In order to satisfy the four dimensions---decentralized, private, hostless, and computable---of an NFP, the implementation has the following requirements. First, minting of a game NFP can be done via CLI or a public website that requires payment in SCRT to the NFP smart contract. The newly minted NFP is then governed by the NFT ownership model: the buyer can be verified as the exclusive owner of the NFP and transferring is done via contract execution. All asset contents (including SVG data) are stored on-chain. Upon minting, the owner of the NFP can then download the SVG to their local device. For each NFP that is minted, its SVG image is unique. When opened in a browser it will render an animated splash page for the game (Active Preview) and a button will appear to connect to the chain and load the game package. When the package is received, the game loads with a list of open matches that have been started by other NFP owners which the player can join, as well as an option to initiate a new match and wait for another player to join. When initiating a new match, an optional wager in SCRT can also be sent to the contract, which must also be met by any joining player. Once joining (or another player joining the owner's game), a new screen appears for both players to set up the game board. Invalid board configurations are rejected by the backend contract. After both players have set up their boards, the match turns begin (with one player randomly chosen to start). Players take turns attacking a cell in the 10x10 grid and the result (a hit or miss) is returned. Vehicle types are only revealed once they are fully destroyed. Turns are enforced by the backend contract and out-of-turn submissions are rejected. Whenever an attack is made, in addition to the hit or miss for the attacker, the opponent is also notified of the event, updating their game state. Once all vehicles are destroyed for one player the match ends and any wager yield is sent to the winner. At any point during play, the state of an open match persists across client restarts and can be recovered by simply reopening the SVG in the browser. 

The game application is 1) decentralized---game state is handled by smart contracts running on Secret Network; 2) private---global game state is only known to the confidential contract, allowing for the implementation of a Bayesian game; 3) hostless---the game can be run by the owner (and only by the owner) simply opening an SVG file in the browser; and 4) computable---the game has an interactive interface with multiple views using HTML5 Web graphics.

\subsection{Implementation}


Here we briefly describe how our game implements all of the features mentioned in Section \ref{methods}. The NFP is first minted by the user on a public website or by CLI, granting them exclusive ownership over a new, unique, private digital asset. The SVG file is then automatically downloaded to their device where they can open it directly in their browser. The SVG contains XML that configures a set of failover API endpoints. The SVG's bootloader fetches the \textit{latest} \texttt{app.js} package from the contract and injects it into the document, transforming into a web application. The now functioning \Client{} then subscribes to a private notifications message channel which is how it will receive updates to the player's game state. A custom designed hot wallet UI component guides the user through requesting a Fee Grant and authorizing the hot wallet account as a token delegate. Once approved, the user can start a new match or browse open matches to join in the lobby. Once a game has started, players configure and submit their grid setup to the contract and begin exchanging attacks one at a time, each execution being signed automatically by the hot wallet. A monorepo containing code to build the SVG template, the source for the game packages, and smart contract backend is available here: \url{https://github.com/nfps-dev/nfp-examples}. Figures~\ref{fig:active-preview}--\ref{fig:game-play} in the Appendix show screenshots of the implemented game.

\subsection{Discussion}

The evaluation game shows that NFPs are a workable model for building decentralized, private, hostless Web applications that can easily match the functionality of a traditional HTML5 Web 2.0 game. The development of design patterns and tooling for NFPs will be important to not only facilitate development (beyond what we've discussed in Section~\ref{sdk}) but also to better identify best practices for managing privacy across the front and backend of the application. This is not a trivial endeavor as each application will have unique privacy guarantee requirements. Furthermore, it should be noted that the degree of decentralization of an NFP depends on the network it is running on. For an application like the one detailed here, a validator set of 75 (Secret Network's active set as of March 2024) is sufficiently decentralized for all intents and purposes, but there are other cases where this might be considered inadequate.

The size of the applications that can be stored on chain remains a limiting factor. At time of writing, the block gas limit on Secret Network is set at 6 million gas units, which enforces an upper bound on the amount of data that can be stored on chain in a single transaction. In our experiments we found that gas used to upload packages scales linearly with the size of package, reaching a current limit of 320KB after compression.
While queries do not cost gas, they are still metered, but since reading is much less expensive than writing, any package written to chain will be accessible to download. The size limit places a restriction on code that can fit in a single package, but with optimization and compression enough leeway exists to build interactive web applications, such as the game described above. If required, the package manager could be extended to allow chunked packages to be downloaded over multiple query requests.


The minted SVG file was 23 KB. Our primary application bundle which gets loaded from the package manager was 338KB before compression -- $ 60\% $ of that going to ambitiously high-resolution graphics -- and 227KB after gzip compression. Keep in mind, the bundle includes all game assets, client and library logic, styling, DOM structure, and UI text for the application. 




This proof of concept demonstrates the basic utility of the NFP model. Some potential applications we envision for NFPs include:

\begin{itemize}
    \item Interactive digital art and graphics which access high-end browser-runnable code such as shaders and web assembly.
    \item Games that reveal encrypted content based on program state and input events.
    \item Scalable, trustless, and decentralized eSports and competitive multiplayer gaming with cheat resistance.  
    \item Censorship-resistant digital content sharing/distribution and Web3 clients.
    \item Applications, such as task workers, with embedded wallets to automate interaction with smart contracts on (any) blockchain. 
    \item Autonomous private overlay networks, such as private peer-to-peer communication networks authenticated via NFP or authenticated access to private oracle services.
\end{itemize}

\section{Conclusion}

In this paper we presented the NFP model for building full-stack Web3 decentralized applications. NFPs leverage existing Web standards and APIs with recent innovations in NFTs and confidential smart contracts to build hostless web applications that encapsulate the usability and interaction of modern Web 2.0 applications joined to a trustless, privacy-preserving backend database service. We demonstrated the utility of the model with an interactive, cheat-resistant and openly verifiable Bayesian game implemented as an NFP. Our next step will be to explore sandbox environments that can safely run NFPs that require access to full Web APIs.

\bibliographystyle{ledgerbib}
\bibliography{nfp}

\newpage

\appendix

\setcounter{section}{0}

\section*{Evaluation game screenshots}

\begin{figure*}[!htp]
  \centering
  \includegraphics[width=.9\linewidth]{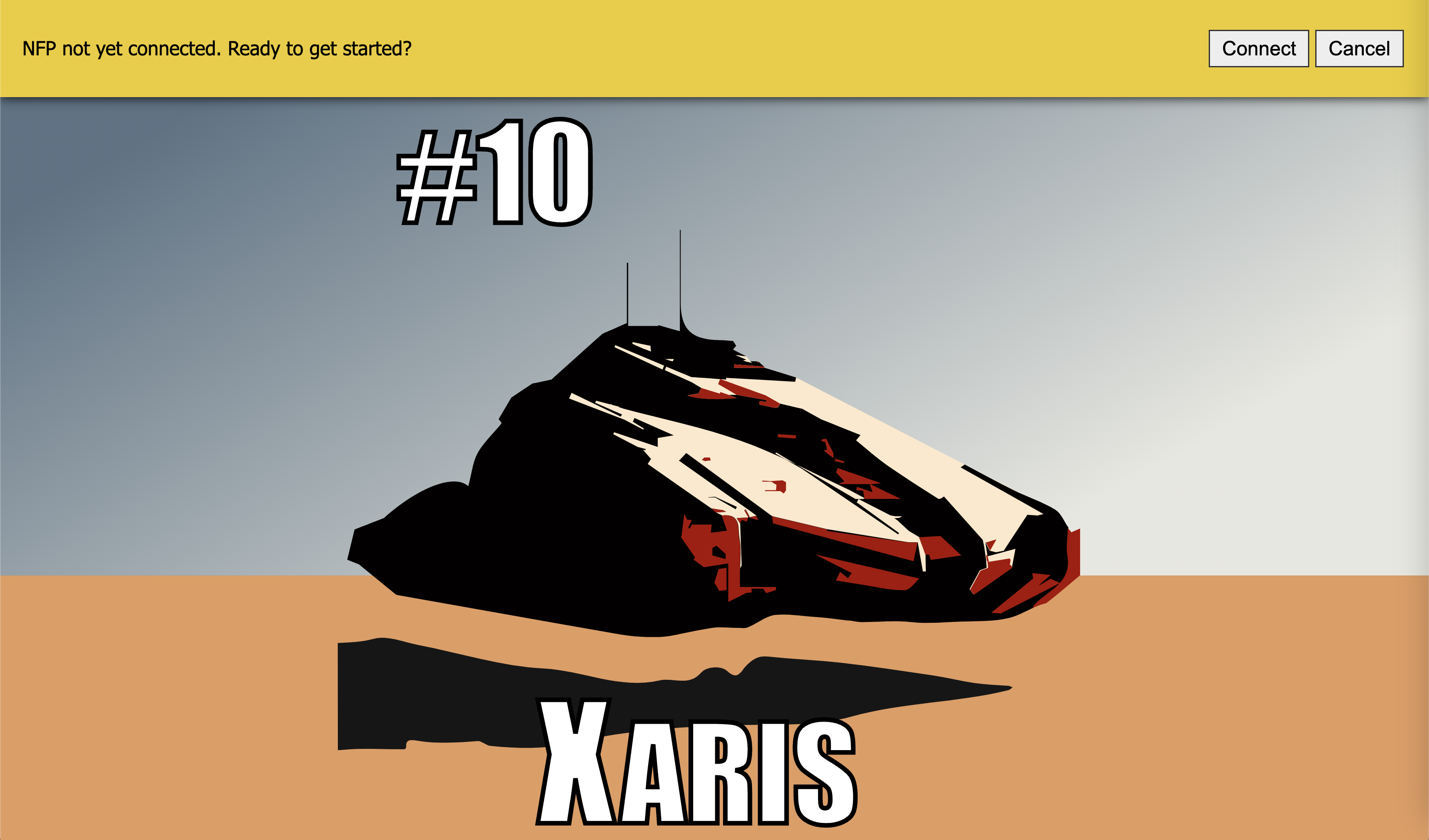}
  \caption{Screenshot of the active preview of the evaluation game NFP. This view is shown when the SVG file is loaded directly from the file system. The ship is animated to float over the desert floor, and the action bar at the top provides the user the ability to connect to the chain and run the game package stored on-chain in the NFP smart contract.}
  \label{fig:active-preview}

\bigskip

  \includegraphics[width=.9\linewidth]{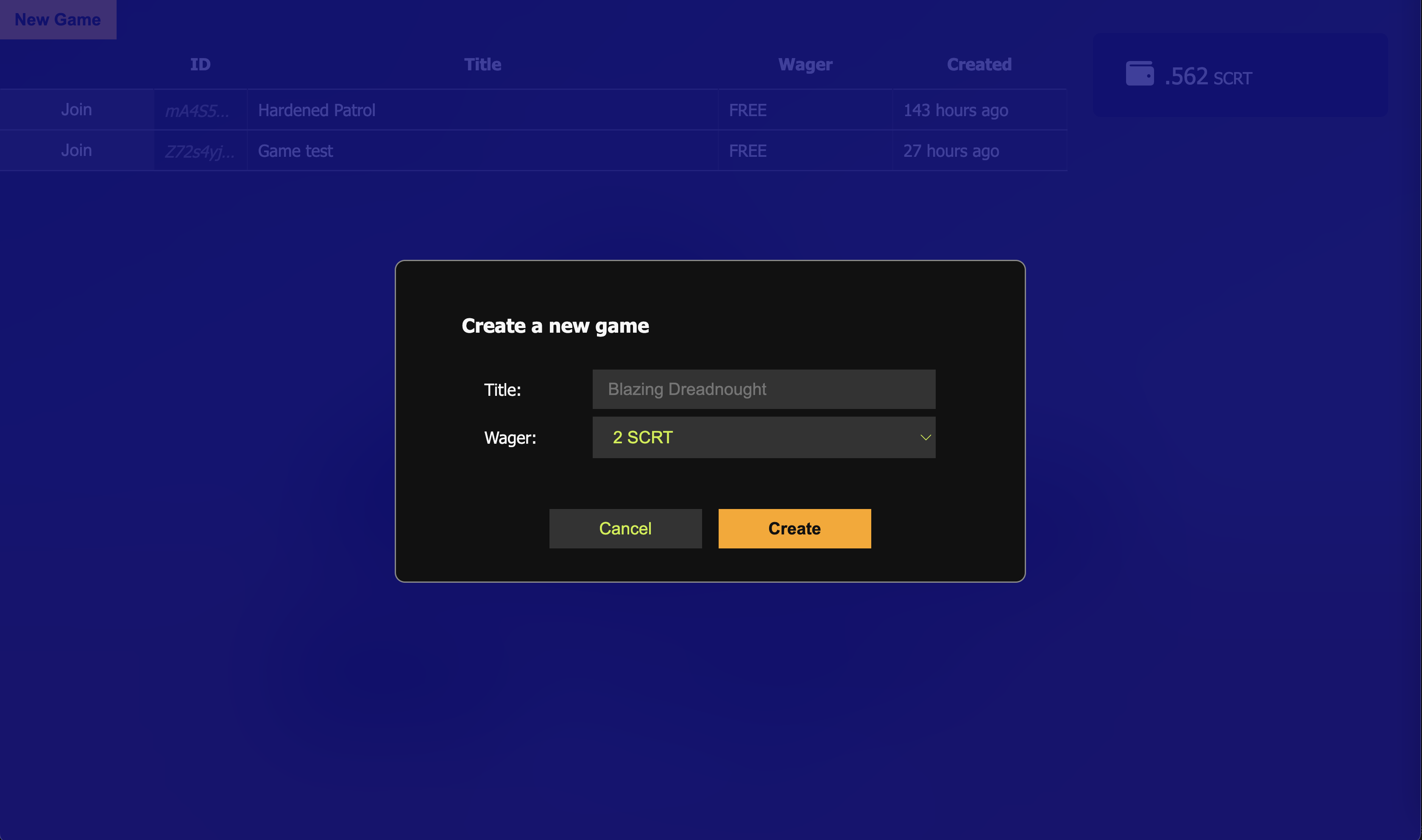}
  \caption{Screenshot of creating a new game after connecting and running the on-chain package code. Behind the dialog overlay, the menu for selecting active games started by other NFP owners is visible.}
  \label{fig:create-game}
\end{figure*}

\begin{figure*}[h]
  \includegraphics[width=.92\linewidth]{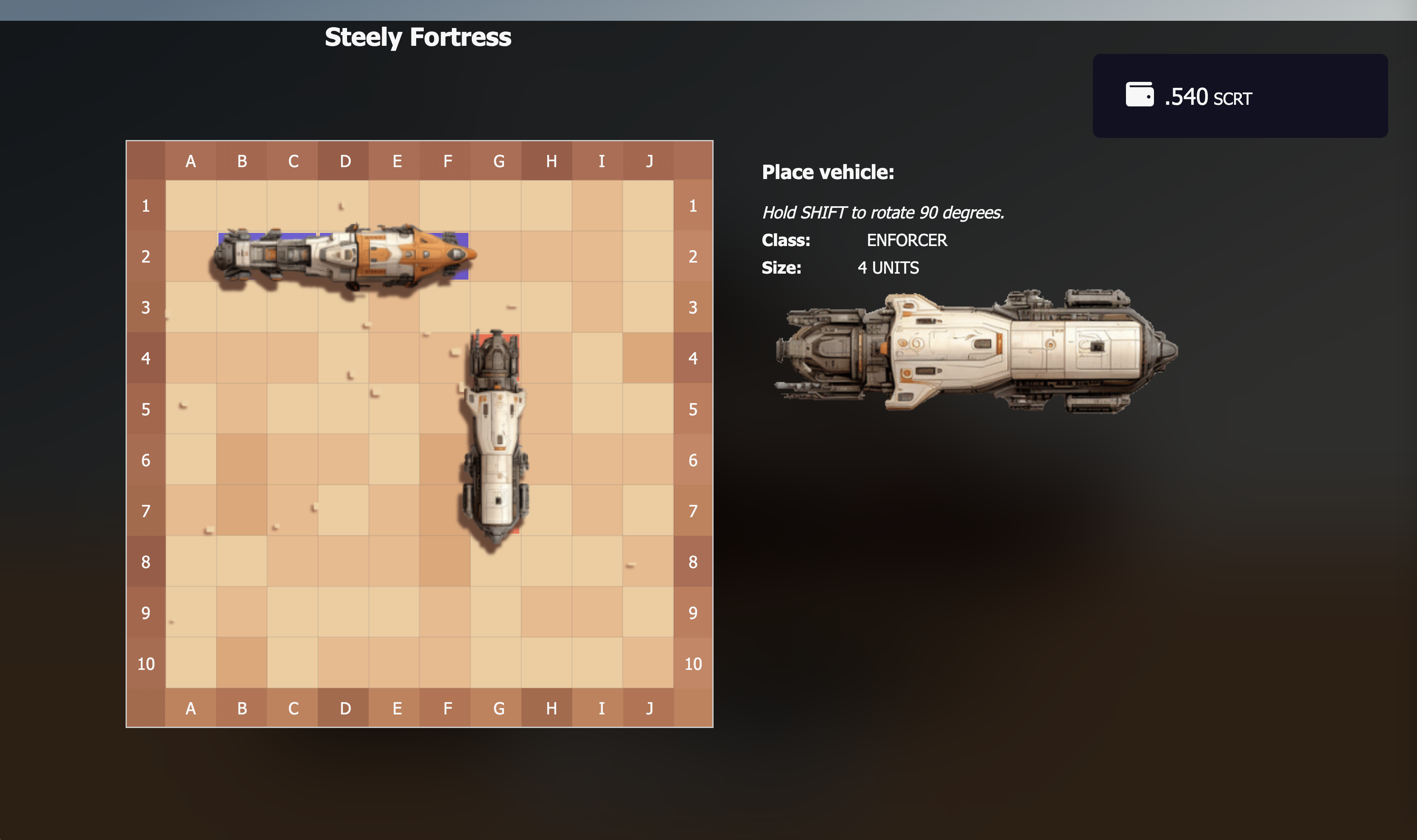}
  \caption{Screenshot of the board setup page for a new game where the player configures the placements of vehicles.}
  \label{fig:board-setup}

\bigskip

  \includegraphics[width=.92\linewidth]{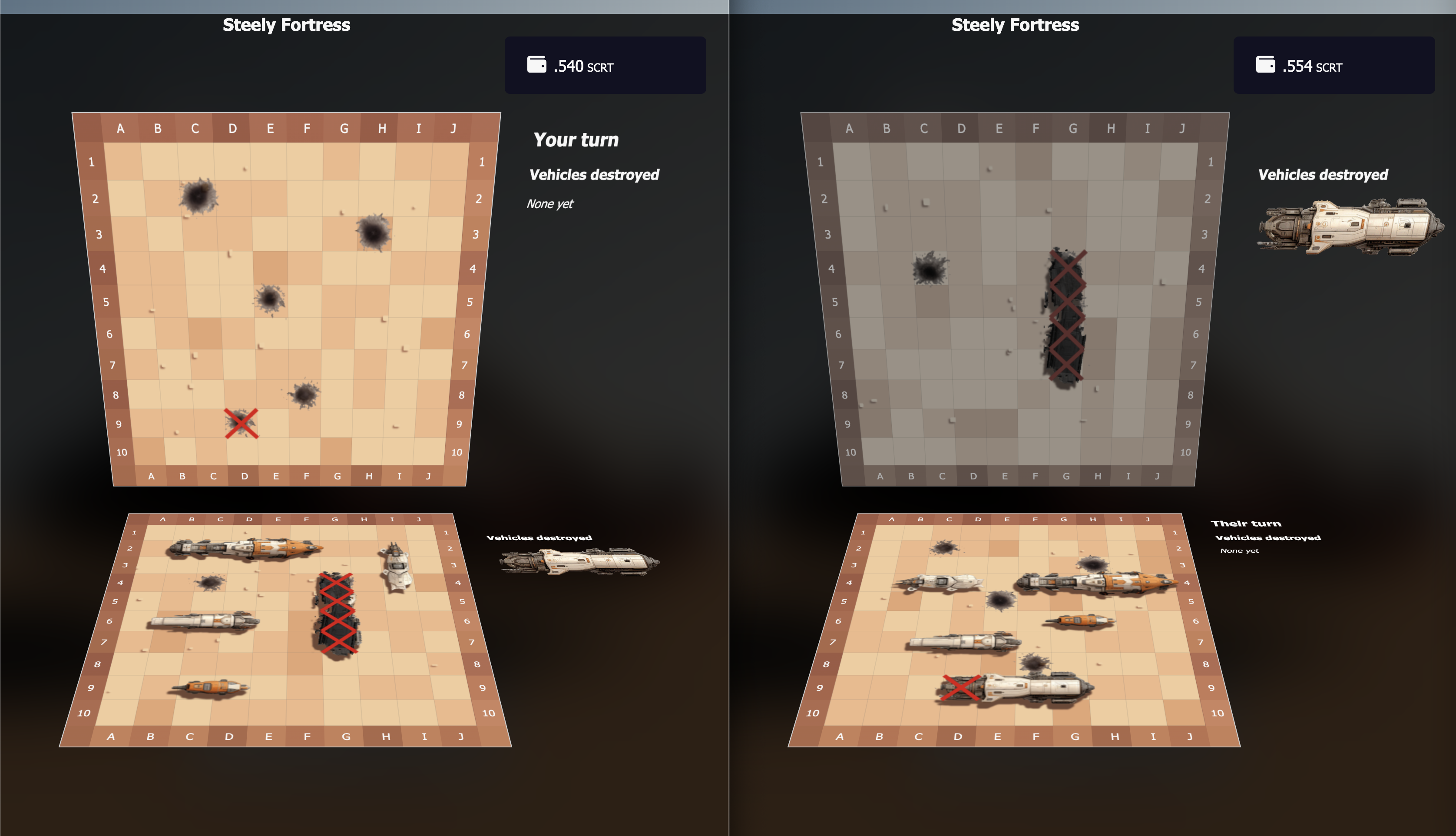}
  \caption{Screenshot of two competing players' screens shown side-by-side as one selects their next attack. The board is interactive with cell selection based on mouse over and click. The state of the game from the player's perspective is shown on separate `home' and `away' boards.}
  \label{fig:game-play}

\end{figure*}





\end{document}